\documentclass[twocolumn,prd,aps,superscriptaddress]{revtex4}

\usepackage{aas_macros}
\usepackage{dcolumn}
\usepackage{bm}
\usepackage{amsmath,amssymb}
\usepackage{color}
\usepackage{graphicx}
\usepackage{url}

\newcommand{\red}[1]{\textcolor{red}{{#1}}}

\def\etal{{\frenchspacing\it et al.}}

\newcommand{\be}{\begin{equation}}
\newcommand{\ee}{\end{equation}}
\newcommand{\ba}{\begin{eqnarray}}
\newcommand{\ea}{\end{eqnarray}}

\newcommand{\Mpc}{\ensuremath{\,{\rm Mpc}}}

\begin{document}
\title{Cosmological test of an extended quintessence model}

\author{Gong Cheng}
\affiliation{National Astronomical Observatories, Chinese Academy of Sciences,
20A Datun Road, Beijing 100101, China}
\affiliation{School of Astronomy and Space Science, University of Chinese Academy of Sciences, Beijing 100049, China}

\author{Fengquan Wu}
\affiliation{National Astronomical Observatories, Chinese Academy of Sciences,
20A Datun Road, Beijing 100101, China}

\author{Xuelei Chen}
\affiliation{National Astronomical Observatories, Chinese Academy of Sciences,
20A Datun Road, Beijing 100101, China}
\affiliation{School of Astronomy and Space Science, University of Chinese Academy of Sciences, Beijing 100049, China}
\affiliation{Department of Physics, College of Sciences, Northeastern University, Shenyang 110819, China}
\affiliation{Center for High Energy Physics, Peking University, Beijing 100871, China}

\date{\today}

\begin{abstract}
 We investigate the cosmological observational test of the extended quintessence model, i.e. a scalar-tensor gravity model 
 with a scalar field potential serving as dark energy,  by using the Planck 2018 cosmic microwave background 
(CMB) data, together with the baryon acoustic oscillations (BAO)  and redshift-space distortion (RSD) data.
 As an example, we consider the model with a Brans-Dicke kinetic term $\frac{\omega(\phi)}{\phi} 
\phi_{;\mu}  \phi^{;\mu} $ and a quadratic scalar potential 
$V (\phi) = A + B (\phi - \phi_0) + \frac{C}{2} (\phi - \phi_0)^2$, which reduces to general relativity (GR) in the  limit 
$\omega(\phi) \to \infty$, and the cosmological constant in the limit $B=C=0$.  In such a model the scalar field typically 
 rolls down the potential and oscillates around the minimum of $V (\phi)$.
 We find that the model parameter estimate for the CMB+BAO+RSD data set is given by
  $\lg \alpha = -3.6 _{-0.54}^{+0.66}~ (68\%)$ ($\alpha$ is defined in Eq.~\ref{eq:omega1}), 
 corresponding to $ 3.8 \times 10^5 < \omega_0 < 9.5 \times 10^7~ (68\%)$,  and $\lg C = 4.9 \pm 1.4~ (68\%) $. 
 However, the GR $\Lambda$CDM model can fit the data almost as good as this extended quintessence model, and is favored by  
 the Akaike information criterion (AIC). The variation of the gravitational constant since the epoch of Recombination is constrained to be
 $0.97 < G_{\rm rec}/G_0 < 1.03~ (1 \sigma)$. In light of recent report that the CMB data favors a closed universe,  we consider the case with non-flat geometry in our fit,  and find that the mean value of $\Omega_k$ shifts a little bit from $-0.049$ to $-0.036$,
 and the parameters in our model are not degenerate with $\Omega_k$. 
 \end{abstract}


\maketitle

\section{INTRODUCTION} 
The scalar-tensor theories of gravity are extended relativistic theories of gravity. 
They can arise naturally as effective theories of the higher 
dimensional theories, for instance, Kaluza-Klein theory,  on four-dimensional spacetime \cite{Clifton2012}. 
They also provide a natural and simple framework 
to model the time variation of the gravitational constant via the dynamics of scalar field \cite{Ooba2016}. 
In the scalar-tensor theory, the Ricci scalar couples to a scalar field. In its simplest version the Brans-Dicke 
theory \cite{Brans1961},  a constant parameter $\omega$ is introduced. In the more general cases, 
$\omega$ can evolve with the Brans-Dicke field $\phi$ and the potential of the Brans-Dicke field 
$V(\phi)$ should be considered. Such a scalar field may also be regarded as an extended 
quintessence model \cite{Perrotta2000, Chen2001}, i.e. a canonical scalar field which couples 
to gravity non-minimally, and can be used to explain the late-time cosmic acceleration \cite{Tsujikawa2013}. 

The action of the extended quintessence model can be written as
\be \label{eq:extended}
S = \int {\rm d}^4 x \sqrt{-g} \left[ \frac{1}{2}F(Q) R - \frac{1}{2} Q^{;\mu} Q _{;\mu} - \widetilde V(Q) + L_{\rm fluid} \right],
\ee
where $F(Q) = \cfrac{1}{8 \pi G_0} + \xi(Q^2 - Q^2_0)$, and $Q_{;\mu}$ denotes covariant derivative of $Q$. 
Rewriting it in the Brans-Dicke form,  $\phi \equiv 8 \pi G_0 F(Q) $, 
and the transformed action is
\ba \label{eq:action}
\nonumber S &=& \frac{1}{16 \pi G_0} \int {\rm d}^4 x  \sqrt{-g}\left[ \phi R -\frac{\omega(\phi)}{\phi} 
\phi_{;\mu} \phi^{;\mu} -  V(\phi) \right]\\
&& + S_{\rm m} (\Psi , g_{\mu \nu}),
\ea
where $\Psi$ denotes the matter field. The dimensionless scalar field $\phi$ has a  present day value $\phi_0$ which 
is very close to $1$,  and its potential $V(\phi)$ is tightly constrained by observations. 
As $\omega(\phi) \rightarrow \infty$, the model reduces to general relativity (GR).
As in the harmonic attractor model\cite{Ooba2016, Ooba2017}, we parameterize $\omega(\phi)$ as
\be \label{eq:omega1}
2 \omega(\phi) +3 = \cfrac{1}{\alpha^2 - \beta \ln(\phi/\phi_0)},
\ee
where $\alpha$ and $\beta$ are model parameters. Note that at present day, $\phi \to \phi_0$, 
the model reduces to the GR case in the limits
$\alpha \rightarrow 0, ~\beta \rightarrow 0$.

Here we consider a model with an evolving scalar field potential  $V(\phi)$ which is 
responsible for dark energy. We assume that the effective potential 
 $V(\phi)$ can be expanded at the low orders as
\be
V (\phi) = A + B (\phi - \phi_0) + \frac{C}{2} (\phi - \phi_0)^2.
\ee
In the limit $B \rightarrow 0, ~C \rightarrow 0$ 
the quintessence reduces to cosmological constant with $A = \Lambda$.

The scalar tensor theory has been tested extensively with various astronomical observations. 
Solar system experiments has put strong constraints on Brans-Dicke theory up to 
$\omega > 40000$ at $2 \sigma$ level \cite{Bertotti2003}. Nonetheless, it 
is conceivable that gravity theory differs from GR in the early universe while it behaves like GR at present. 
It is therefore necessary to probe the behavior of gravity in different environments and 
scales, the cosmic evolution can provide a good laboratory to test gravity in the low 
density and low curvature regime, and the CMB data and large 
scale structure (LSS) data can be used to constrain such evolution \cite{Chen1999,Wu2010-1}. 
While CMB provides the cleanest observational data, the geometrical redshift-distance relations measured by 
galaxy redshift surveys are useful for breaking the degeneracies in the CMB data. For the LSS, the simplest 
approach is to use the baryon acoustic oscillation (BAO) distance measurements, which we will use in the present paper, 
though there are also other  approaches,  for example measurement based on topology \cite{2012ApJ...747...48W}.
The redshift-distance relation as measured by 
standard candles such as the Type Ia supernovae explosion is less reliable for this test, as the 
luminosity of the supernovae may depend on the  gravitational constant, which varies in the scalar-tensor model, though such 
evolution has not been detected observationally \cite{Li_2015}. In the present work, we will only use the distance measured with 
the CMB and LSS.   

In addition, the growth of structure is 
expected to be suppressed or enhanced in modified gravity compared to the standard GR model. 
So complementing distance measurements, the growth function is particularly sensitive to impose constraints on the 
modified gravity models \cite{Huterer2015}. 

The CMB data together with LSS data had been used to 
constrain scalar-tensor models. In Ref. \cite{Wu2010-2}, 
the region of $-120.0 < \omega < 97.8$ was excluded 
at $2\sigma$ level by using WMAP 5 year data, other CMB experiments data and 
LSS data measured by luminous red galaxy (LRG) survey of Sloan 
Digital Sky Survey (SDSS) data release 4. The constraint was improved 
to $ \omega < -407.0$ or $\omega > 175.87$ by using {\it Planck} data \cite{Li2013}. 
Avilez and Skordis \cite{Avilez2014}  reported $\omega > 1808$,  by using {\it Planck} temperature and WMAP 9-year polarization data.
Ooba \etal ~\cite{Ooba2016, Ooba2017} obtained constraints on the harmonic attractor model, and some more recent  constraints were given in \cite{Umilt__2015,Ballardini_2016,Ballardini_2020,Rossi_2019}. In addition to the Brans-Dicke gravity, there have also been many 
investigations on more general classes of scalar-tensor models, 
such as the $f(R)$ gravity, early modified gravity \cite{EMD1, EMD2}, and the Horndeski gravity which is the most general form with second-order 
field equations; see Refs. \cite{Martino2015, Ishak2018} for reviews. 
Note that these limits are dependent on the prior used, even for the same data, quite different limits can be derived with different priors.

The commence of multi-messenger astronomy with gravitational wave (GW)  detection opens up a new window to test gravity theories. 
Based on the detections of GW signal produced by the binary neutron star merger (GW170817) \cite{Abbott2017-1} 
and its electromagnetic counterpart (gamma ray burst GRB170817A) \cite{Abbott2017-2, Abbott2017-3}, the speed of GW is 
constrained to be $c_{\rm T}^2 -1 \lesssim 10^{-15}$. These measurements have several crucial implications for cosmological 
scalar-tensor theories \cite{Ezquiaga2017, Sakstein2017, Creminelli2017, Baker2017, 2018PDU....22..108C, 2019CQGra..36a7001C}. 
As a result, the quartic and quintic Galileons 
are strongly excluded. The remaining viable models include simple Horndeski, e.g., Brans-Dicke and $f(R)$, and specific 
models beyond Horndeski theory which is either conformally equivalent to theories with $c_{\rm T} = c$ or disformally fine-tuned.

Due to the vast number of gravity models in the market, it is important to test gravity in a model-independent way. 
In the past few years, several perturbation parameterizations are proposed and implemented in the Boltzmann solvers 
{\tt CAMB} \cite{2000ApJ...538..473L, 2002PhRvD..66j3511L}, such as {\tt EFTCAMB} \cite{Hu2014} and {\tt MGCAMB} \cite{Hojjati2011}. 
Although these formalisms cover a broad class of models, they have some limitations. There are a large number of free functions 
to be constrained and simple parameterizations of these functions are unlikely to be successful, even missing the signature of 
modified gravity in the observations. Also, the connection between the formalism and physical models is not intuitive 
\cite{Linder2017, Planck2016-14}.

Most previous works on the scalar-tensor theory were performed for cosmological model of flat geometry. 
Recently, it has been shown that the {\it Planck} CMB power spectra prefer a closed universe if the CMB lensing data is not 
used \cite{2020NatAs...4..196D} for the models with gravity given by general relativity. The preference for the closed universe 
exists in both {\it Planck} 2015 and {\it Planck} 2018 data release 
and can not be removed by switching the likelihoods \cite{2020MNRAS.496L..91E}. The positive curvature can explain the 
anomalous lensing amplitude $A_{\rm lens}$ but lead to the discordances for the local observations such as BAO. These discordances may 
originate from systematics or new physics. 
In Ref. \cite{2021ApJ...908...84V}, the constraint from Planck data and cosmic chronometers (CCs) gives $\Omega_k = -0.0054 \pm 0.0055$,
without causing strong tension between Planck and CC data in the non-flat universe.

To constrain the model, we use the latest cosmological observations, including the CMB data together with BAO data and RSD data. 
We also investigate the curvature in this model to explore the possibility that the CMB temperature and polarization data 
might be compatible with the flat universe in the context of modified gravity, so that the discordances caused by the closed universe 
can be eliminated.

The rest of the paper is organized as follows: in Sec. II, we discuss the formalism of our computation, including background (II.A) and perturbation evolution (II.B), as well as numerical method (II.C). In Sec. III we review the observational data used. We present our results in Sec. IV, and  conclude in Sev. V. 

In this paper we use natural units with $c=1$. Following the convention of the {\tt CLASS} code, 
all physical quantities are in unit of ${\rm Mpc}^n$, so for example, the units of 
$V(\phi), A, B, C, \rho, p$ are all ${\rm Mpc^{-2}}$.

\section{Formalism}

The generalized Einstein equation and the equation of motion for the scalar field in this model are
\ba
\nonumber \phi G_{\mu \nu} &+& \left[\Box \phi + \frac{1}{2} \frac{\omega(\phi)}{\phi} (\nabla \phi)^2 + \frac{1}{2} V(\phi) \right] 
g_{\mu \nu} -\nabla_{\mu}\nabla_{\nu} \phi \\
&-& \frac{\omega(\phi)}{\phi} \nabla_{\mu}\phi \nabla_{\nu}\phi = 8 \pi G_0 T_{\mu \nu}, \\
\nonumber [ 2 \omega(\phi) &+& 3] \Box \phi + \frac{{\rm d} \omega(\phi)}{{\rm d} \phi} (\nabla \phi)^2 + 2 V(\phi) 
- \phi \frac{{\rm d} V(\phi)}{{\rm d} \phi} \\
&=& 8 \pi G_0 T,
\ea
where $T$ is the trace of the energy-momentum tensor.
We use a perturbative approach to compute the observables in this model.

\begin{figure*}
\begin{center}
\begin{tabular}{cc}
     \includegraphics[width=0.37\textwidth]{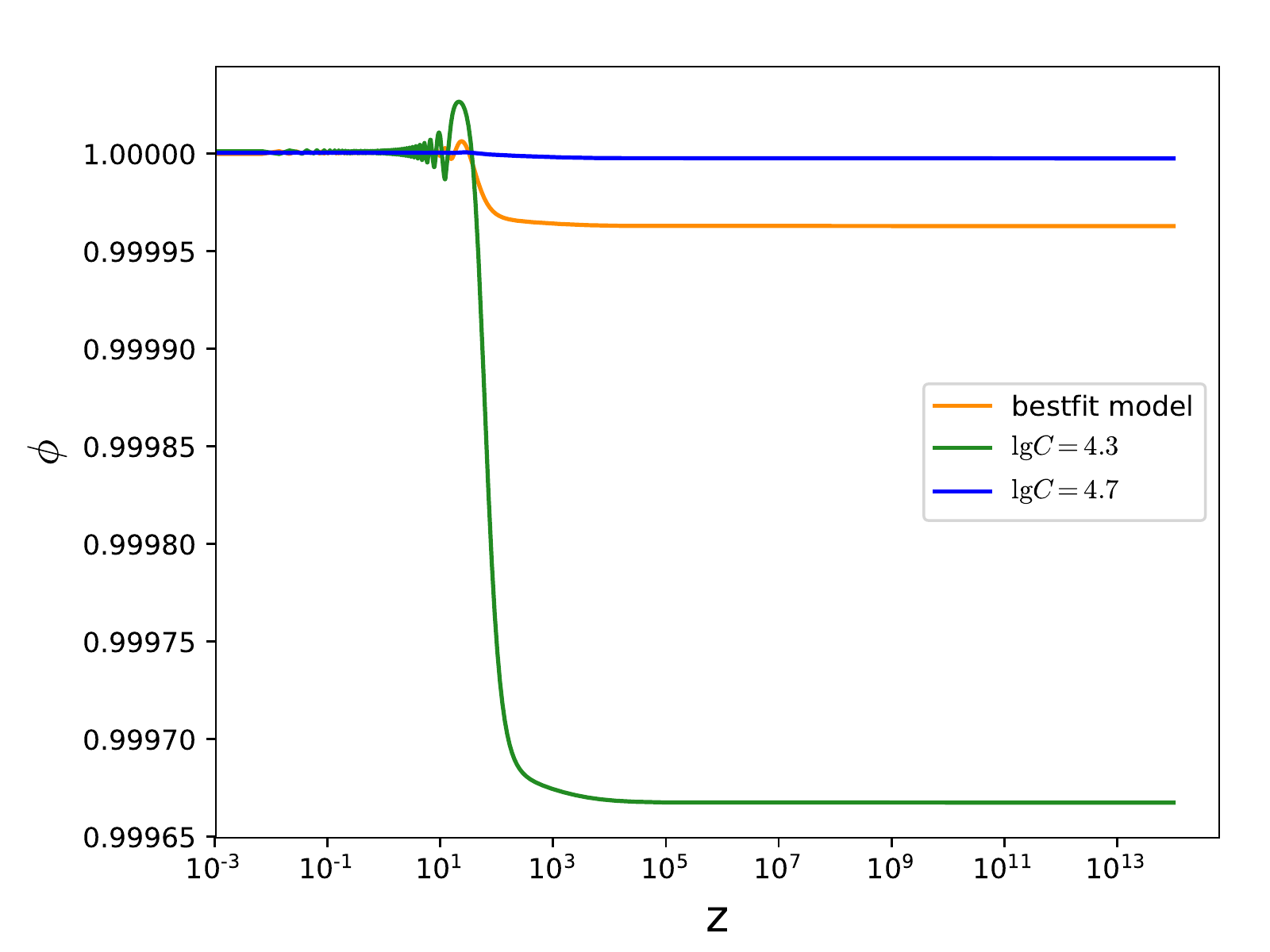}
     \includegraphics[width=0.37\textwidth]{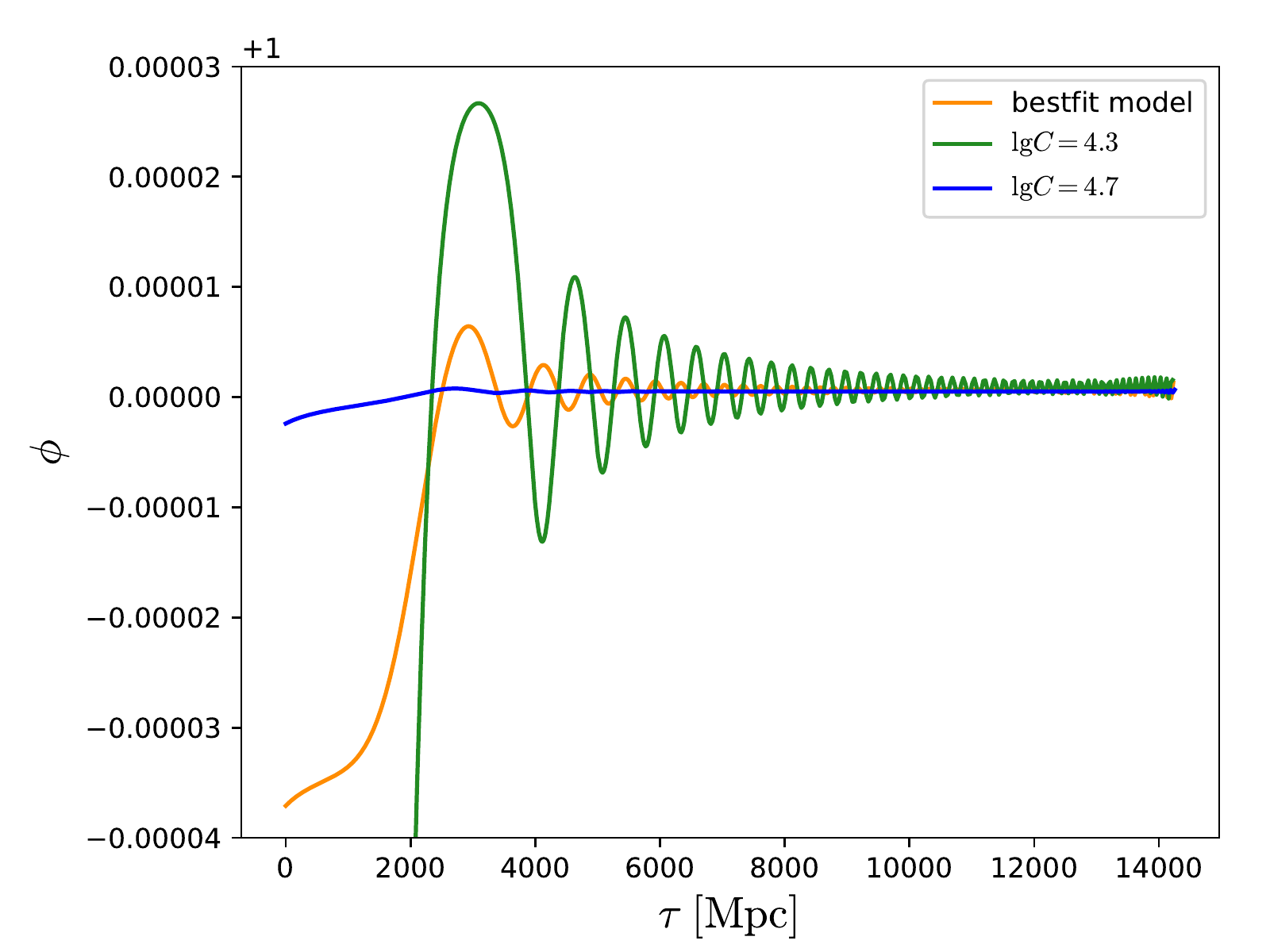}\\
     \includegraphics[width=0.37\textwidth]{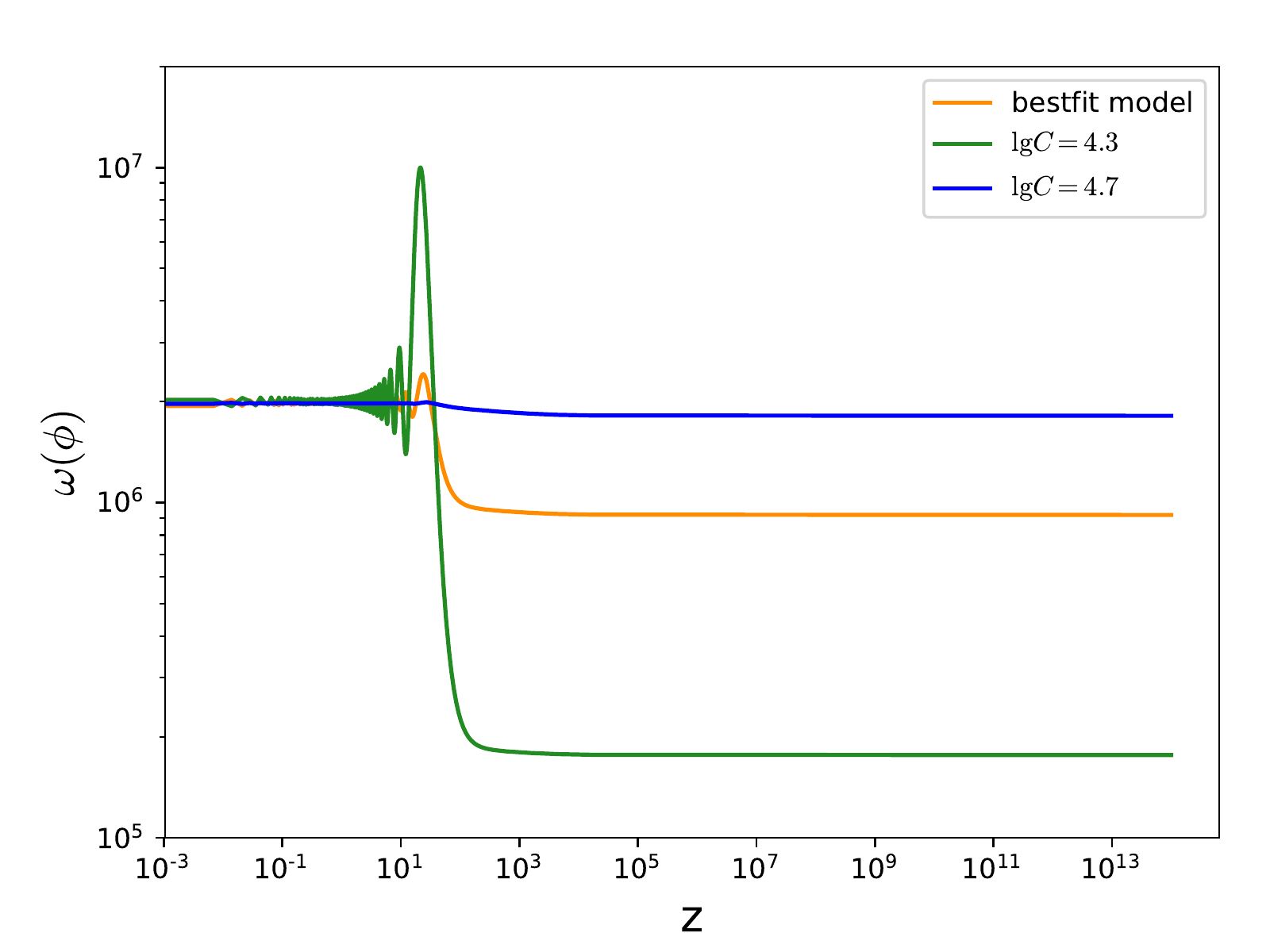}
     \includegraphics[width=0.37\textwidth]{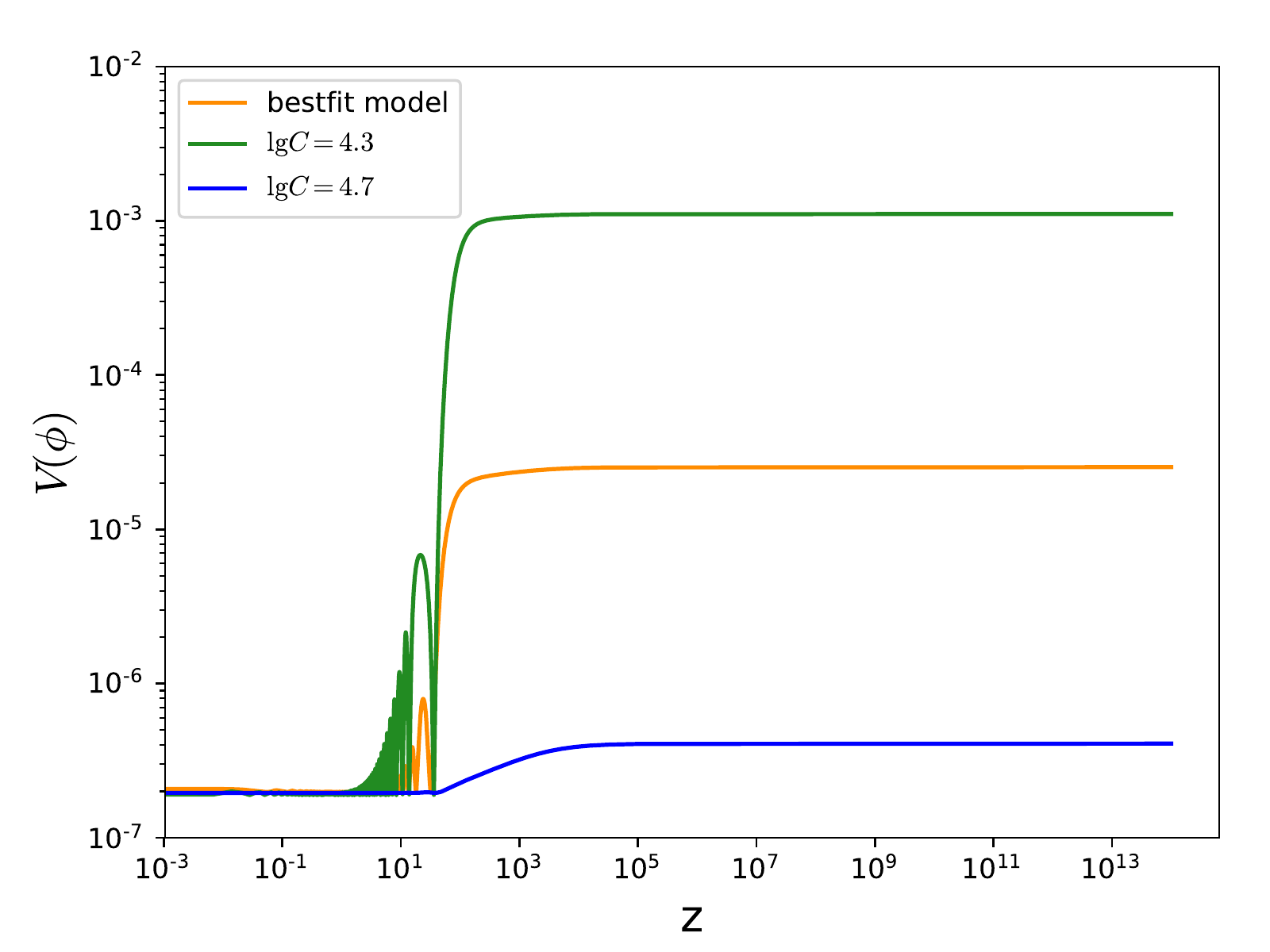}\\
\end{tabular}
\caption{Top Left: The redshift evolution of the scalar field $\phi$ in a few of our extended quintessence models.
The orange solid line represents the best-fit model with the CMB+BAO+RSD data set. The green and blue solid lines show models in which 
$\lg C$ is set to some selected values within the $1 \sigma$ bound, while other parameters are set to the best-fit values. Top Right:
Evolution of $\phi$ with respect to the conformal time $\tau$. Bottom Left:
Evolutions of the generalized Brans-Dicke parameter $\omega (\phi)$. Bottom Right: Evolution of the scalar potential $V (\phi)$.}
\label{fig:back}
\end{center}
\end{figure*}

\subsection{Background Evolution}
The equations governing the homogeneous 
background evolution of the universe are
\ba
\nonumber H &=& \frac{1}{a} \left[-\frac{{\phi}'}{2 \phi} + \sqrt{\frac{a^2 V}{6\phi} + \frac{a^2\rho}{\phi} 
+ \frac{2\omega+3}{12} \left(\frac{{\phi}'}{\phi}\right)^2 
-\kappa } \right], \label{eq:friedmann}\\
&& \\
\nonumber {H}' &=& -\frac{3 a p}{2 \phi} -\frac{3}{2} \cfrac{{a}'^2}{a^3} - \frac{\kappa}{2 a} - \frac{{\phi}''}{2 a \phi} 
- \cfrac{{a}'}{a^2} \frac{{\phi}'}{\phi} - \frac{\omega {\phi}'^2}{4 a \phi^2} + \frac{aV}{4 \phi}, \\
&& \\
\nonumber {\phi}'' &=& -2  \cfrac{{a}'}{a} {\phi}' + \frac{1}{2\omega +3} \times \\
&& \left[3 a^2 (\rho - 3 p) + 2 a^2 V - a^2 \phi \frac{{\rm d} V}{{\rm d} \phi} 
- {\phi}'^2 \frac{{\rm d} \omega}{{\rm d} \phi} \right],
\label{eq:phi}
\ea
where the prime denotes the derivative with respect to the conformal time $\tau \equiv \int {\rm d} t/a(t)$, and 
$H \equiv \cfrac{{a}'}{a^2}$ is the proper time Hubble rate, and
$\kappa = -\Omega_k H_0^2$ is a number  characterizing the spacetime curvature in the cosmological
model, $\rho$ and $p$ are the rescaled total density and pressure of all matter components,
$$\rho = \cfrac{8 \pi G_0}{3} \rho_{\rm physical}, ~~ p = \cfrac{8 \pi G_0}{3} p_{\rm physical}.$$

The effective gravitational constant measured by Cavendish-type experiments is given by \cite{Brans1961}
\be
G(\phi) = \frac{G_0}{\phi} \frac{2\omega(\phi) + 4}{2\omega(\phi) + 3}.
\ee
so the value of $\phi$ at present day is set to be 
\be \label{eq:present-day}
\phi_{0} = \frac{2 \omega_0 + 4}{2 \omega_0 + 3}.
\ee

Fig. \ref{fig:back} shows the evolution of the scalar field $\phi$ for three models: 
 the best-fit model obtained with the latest cosmological observations as detailed in the later sections, and the models 
with parameters $\lg C = 4.3$ and $\lg C = 4.7$. As we shall see later, all of these models are  
within the $1 \sigma$ bound of $\lg C$.

The field $\phi$ began to increase after matter-radiation equality and converged to the present-day $\phi_0$. 
The corresponding evolutions of the generalized Brans-Dicke parameter $\omega (\phi)$ and the potential term $V (\phi)$ 
are plotted in the bottom panels of Fig.~\ref{fig:back}.  Note that in all of these models the present day gravitation is 
indistinguishable from GR as  $\omega (\phi) $ are high above the current solar system experimental limit. 
The scalar potential $V (\phi)$ decreases from a higher value to its present day
value for the three example models shown in the figure. Its value at high redshift can be $1 \sim 3$ 
orders of magnitude higher than the present day, so in these models the dark energy can be important 
in the early evolution of universe.

In all the subplots of Fig. \ref{fig:back}, the curves oscillated before converging to the present day value. It is easy to understand such oscillation behavior: the scalar field $\phi$ eventually settles down at the minimum of the potential, but near the bottom of the potential 
it behaves like a damped oscillator.  
If we focus on the oscillatory terms containing ${\phi}''$, ${\phi}'$, $\phi$ and the dominant constant term, and neglect the other non-oscillatory 
terms of Eq.~(\ref{eq:phi}), the equation of motion reduces to a damped harmonic oscillator form, 
\begin{equation}
\label{eq:general}{\phi}'' +  2 H a {\phi}' +  k(\tau) (\phi - \phi_0) = 0,
\end{equation}
where $ k(\tau) = \frac{a^2}{2 \omega+3} (C\phi_0-B)$.
If we take $H a, k(\tau)$ as slowly varying, this can be solved analytically with the Wentzel–Kramers–Brillouin (WKB) approximation (see e.g. \cite{mathews1970mathematical}).
  To first order,
\begin{eqnarray}
\phi(\tau) - \phi_0 &=& I_0 f(\tau)^{-\frac{1}{4}} e^{{-\int H a ~d\tau}} \nonumber\\
&&\times \cos \left[\int^{\tau}_{\tau_2} (\sqrt{f(x)} - \frac{b(x)'}{4\sqrt{f(x)}} )dx\right]. 
\label{eq:appx1} 
\end{eqnarray}
where $I_0$ is a constant, and 
\ba
f(\tau)  &=& \cfrac{a^2}{2\omega(\phi)+3} (C\phi_0-B) - (aH)^2.
 \ea
 As $\phi_0 \approx 1, B \ll C$, and $(Ha)^2$ is also very small,  these terms are negligible and 
 \ba
f(\tau) &\approx& \cfrac{a^2 C}{2\omega(\phi)+3},\\
&=& a^2 C \left[\alpha^2 - \beta \ln (\phi/\phi_0)\right].
\ea
where the last is obtained by substituting Eq.~(\ref{eq:omega1}).  
Also,
$$
e^{-\int Ha ~d\tau} =e^{-\int \frac{da}{a}} = a^{-1}
$$
 we then have  approximately  
\ba
\nonumber \phi(\tau) - \phi_0 \sim  I_0 a^{-\frac{3}{2}} \left[ C \left(\alpha^2 - \beta \ln (\phi/\phi_0)\right) \right]^{-\frac{1}{4}}\\
\label{eq:final} \times \cos \left[ \sqrt{a^2 C \left(\alpha^2 - \beta \ln (\phi/\phi_0)\right)}\tau + \psi \right].
\ea

Compared with the oscillation term, the Hubble drag term is small, 
$H^2/\left\{C \left[\alpha^2 - \beta \ln (\phi/\phi_0)\right]\right\} \sim 7 \times 10^{-4}$, 
so the oscillation is under-damped. 
To examine the accuracy of this solution, we compare the period and amplitude obtained from the above analytical
solution and the numerical solution.
We take the model $\lg C = 4.3$ (the green line in Fig.~\ref{fig:back}) as an example. The conformal period 
estimated from Eq.~(\ref{eq:final}) at  $z=5$ $(c \tau= 6206 \Mpc)$ is $c T = 537.3  \Mpc$, which agrees well with the $539.6 \Mpc$ as 
measured from the numerical solution.
The ratio of amplitude between first and second peak estimated from the above equation is about \red{4.2}, while
the one obtained from numerical solution is 2.7.

These oscillations occur mostly at high redshifts, and the amplitude for the $\omega(\phi)$ and 
$V(\phi)$ dropped low at low redshift. Nevertheless, they may be observable in high redshift precision observations in the future.

\subsection{Perturbation Evolution}
In perturbation theory, $g_{\mu\nu}= a^2 (\gamma_{\mu\nu}+h_{\mu\nu})$, where $\gamma_{\mu\nu}= \mathrm{diag}(-1, 1, 1, 1)$ 
and $h_{\mu\nu}$ are respectively the unperturbed and perturbed part of the metric. We choose to work in the 
synchronous gauge, where $h_{00} =h_{0i}=0$, and follow the formalisms 
developed in Refs. \cite{Kodama1984, Ma1995, Hu1998, Tram2013}. The perturbations could be decomposed into 
the eigentensors of the Laplacian $\nabla^2 Q^{(m)} = -k^2 Q^{(m)}$, 
\ba
h_{ij} &=& \sum_{m} 2 h_L Q^{(m)} \gamma_{ij} + 2 h_T Q_{ij}^{(m)} \\
&=& \sum_{m} \frac{h}{3} Q^{(m)} \gamma_{ij} - (6\eta + h) Q_{ij}^{(m)}, \\
\delta \phi &=& \sum_{m} \chi^{(m)}  Q^{(m)}.
\ea
The stress energy tensor can be expressed as $T_{\mu\nu} = \bar{T}_{\mu\nu} + \delta T_{\mu\nu}$, and the unperturbed 
components are $\bar{T}^0_{~0} = -\rho$, $\bar{T}^0_{~i} = \bar{T}^i_{~0} = 0$ and $\bar{T}^i_{~j} = p \delta^i_{~j}$. 
The stress energy perturbations can likewise be decomposed as 
\ba
\delta T^0_{~0} &=& - \sum_{m} \delta \rho^{(m)} Q^{(m)}, \\
\delta T^0_{~i} &=& \sum_{m} \frac{(\rho + p) \theta^{(m)}}{k} Q_i^{(m)}, \\
\delta T^i_{~0} &=& - \sum_{m} \frac{(\rho + p) \theta^{(m)}}{k} Q^{(m) i}, \\
\delta T^i_{~j} &=& \sum_{m} \delta p^{(m)} \delta^i_{~j} Q^{(m)} + \frac{3}{2} (\rho + p)\sigma^{(m)} Q^{(m) i}_{~~~~j}.
\ea

If we only consider the scalar perturbations, the perturbed equations read
\ba
{h}' &=& \left[k^2s^2 \eta + \frac{3a^2}{2\phi}\delta \rho - \frac{w{\phi'^2} \chi}{2\phi^3} 
+ \frac{1}{\phi^2}\left(-\frac{3}{2}a^2\rho \chi + \frac{{\phi'^2} {\rm d} \omega}{4 {\rm d} \phi}\chi \right. \right. \nonumber \\
\nonumber &&\left. + \frac{\omega {\phi}' {\chi}'}{2} + \frac{3{a'}}{2a}{\phi}'\chi - \frac{1}{4}a^2V\chi \right) 
+ \frac{1}{\phi}\left(-\frac{1}{2}k^2\chi - \frac{3{a}'}{2 a}{\chi}' \right.\\ 
&&\left. \left. + \frac{a^2 {\rm d} V}{4 {\rm d} \phi} \chi\right)\right] \left(\frac{{a}'}{2a} + \frac{{\phi}'}{4 \phi}\right)^{-1}, \\
 {\eta}' &=& \left\{\frac{3a^2}{2\phi} (\rho + p)\theta + \frac{1}{2}\kappa {h}' + k^2 
\left[\frac{\omega {\phi}'\chi}{2\phi^2} + \frac{1}{2\phi} \bigg({\chi}' \right. \right. \nonumber \\
&& \left. \left.\left. - \frac{{a}'\chi}{a}\right)\right]\right\} \frac{1}{k^2s^2}, \\
{h}'' &=& -\left(2\frac{{a}'}{a} + \frac{{\phi}'}{\phi}\right){h}' + 2k^2s^2\eta - \frac{9a^2}{\phi} 
\delta p + \frac{9a^2\chi}{\phi^2} p \nonumber \\
\nonumber && + \frac{3{\phi'^2}\omega \chi}{\phi^3} + \frac{1}{\phi^2} \left(-\frac{3 {\rm d} \omega}{2 {\rm d} \phi}{\phi'^2}\chi -3\omega{\phi}'{\chi}' 
+ 3 {\phi}'' \chi \right. \\
&& \left. - \frac{3}{2}a^2V\chi + 3\frac{{a}'}{a}{\phi}'\chi \right) + \frac{1}{\phi}\left(-3{\chi}'' - 3\frac{{a}'}{a}{\chi}' \right. \nonumber \\
&& \left. + \frac{3a^2{\rm d} V}{2{\rm d} \phi} \chi - 2k^2\chi \right),
\ea
\ba
{\alpha}' &=& -\left(2\frac{{a}'}{a} + \frac{{\phi}'}{\phi}\right)\alpha + \eta - \frac{9a^2}{2k^2\phi}(\rho + p)\sigma 
- \frac{\chi}{\phi},\\
\nonumber{\chi}'' &=& -2\frac{{a}'}{a}{\chi}' - k^2\chi - \frac{1}{2}{\phi}'{h}' + \frac{1}{2\omega+3} 
\left[-2\frac{{\rm d} \omega}{{\rm d} \phi} \chi \bigg({\phi}'' \right. \\
\nonumber&& + 2\frac{{a}'}{a}{\phi}' \bigg) - \frac{{\rm d}^2 \omega}{{\rm d} \phi^2} {\phi '}^2 \chi - 2 \frac{{\rm d} \omega}{{\rm d} \phi} {\chi}' {\phi}' + a^2\frac{{\rm d} V}{{\rm d} \phi} \chi \\
&& - a^2\frac{{\rm  d}^2 V}{{\rm d} \phi^2} \phi \chi + 3a^2\bigg(\delta \rho - 3 \delta p\bigg) \bigg],
\ea
where $s^2 = 1 - \cfrac{3\kappa}{k^2}$ and $\alpha = ({h}' + 6{\eta}')/2k^2$.

\subsection{Numerical Methods}
We modify the publicly available Boltzmann code {\tt CLASS} to numerically solve 
the equations above and compute CMB temperature and polarization anisotropy. 

To recover the present day value of the effective gravitational constant given by  Eq.~ (\ref{eq:present-day}), we use a ``shooting" algorithm \cite{Chen1999}, i.e. we evolve the 
background evolution equation from a very early time with a set of given model parameters and initial value of 
the scalar field $\phi_i$ to the present day time, $\phi_i$ is adjusted gradually until the Eq.~ (\ref{eq:present-day}) is satisfied to  
the required precision.  In each ``shooting", the model parameters which affect the background dynamics must be specified, these 
include the spatial curvature $\Omega_k$, the total non-relativistic matter density $\Omega_m$, the Hubble constant $H_0$,  
gravity parameters $\alpha, \beta$ and the scalar parameters $A, B, C$, and the second order differential equation 
also requires the initial values of the 
field $\phi_i$ and ${\phi}'_i$. However, as in many quintessence models,  the ${\phi}'$ damps to a ``terminal velocity" at a later
time, so that within a plausible range ${\phi}_i'$ practically does not affect the result. Of course, when the deviation is too large, 
there would be no $\phi_i$ which could yield the desired solution, but the solution could be found if the deviation from GR is within reasonable range. In the simplest Brans-Dicke model, $\phi_0$ increases monotonically with $\phi_i$. However,
in the present model the evolution of $\phi$ is more complicated. As a result, in some parameter space, 
more than one $\phi_i$ can evolve
to the same desired $\phi_0$. For this case, we choose the one which is closest to $\phi_0$. 

Also, at $z=0$, the left hand side (LHS) of Eq.~(\ref{eq:friedmann}) reduces to $H_0$, and if we neglect the kinetic term 
(the first term) in the right hand side (RHS) of Eq.~(\ref{eq:friedmann}), 
which is only of the order $\sim 10^{-6}$ of the second term, we see this determines the value of $V(\phi)$, which in turn determines
almost completely the value of $A$, which must be numerically very close to the cosmological constant in the $\Lambda$CDM model.
We update ${\phi}'_0$ and $A$ by the iteration process in the ``shooting" algorithm.

In order to constrain the parameters in the gravity model with observations, we use the publicly 
available code {\tt Monte Python} \cite{Audren2013, Brinckmann2018}, 
which adopts the Markov Chain Monte Carlo (MCMC) method to explore the parameter space, and obtain Bayesian estimate of the parameters. At each point in the parameter space, the form of the potential $V(\phi)$ must be specified for {\tt CLASS} to make the 
background run. 

The parameters that varied in the Markov chains are the dynamical parameters listed above ($\Omega_k, H_0, \alpha, \beta, B, C$, 
matter physical density $\omega_m = \Omega_m h^2$, where $h$ is defined by $H_0 = 100 h {\rm km} s^{-1} \Mpc^{-1} $), 
and the baryon physical density $\omega_{\rm b} = \Omega_b h^2$, (the cold dark matter physical density is given by 
$\omega_{\rm cdm}=\Omega_{\rm cdm} h^2 = \omega_m - \omega_{\rm b}$), the amplitude of scalar perturbations $\ln10^{10}A_s$, 
the spectral index for scalar perturbations $n_s$, reionization optical depth $\tau_{\rm reio}$.   
Following the {\it Planck} analysis, we assume two massless and one massive
neutrinos with mass $m=0.06 ~{\rm eV}$ in this paper. We have also checked that if assuming three massless neutrinos, 
the results only have minor changes.

To avoid deviating too much in parameter space which could derail the code running,  we set the priors as
\ba
\nonumber \lg \alpha &\in& (-6, ~~-1),\\
\nonumber \beta &\in& (0, ~~1), \\
\nonumber B &\in& (-0.1, ~~0.1), \\
\nonumber \lg C & \in & (-10,20).
\ea
Note that the limits derived with the Bayesian method is  dependent on the priors. The limits derived for priors with the
linear and logarithmic distribution can be different. Here we adopt the logarithmic prior for $\alpha$, this allow us to measure its 
posterior distribution near $\alpha=0$ in more detail. We note that the GR limit $\alpha=0$ corresponds to $\lg \alpha = -\infty$, which can 
not be attained in the current range of parameter, and if the peak of distribution is close to the lower limit of $\lg \alpha$, it could be the  
GR limit is favored. We shall see later that fitting the observation data constrain these parameters to a range well within these bounds. 
In a trial run we found that the data favors a positive $C$, which allow us to also take a logarithmic prior on $C$ to explore a large range 
of its value.

\section{Observational Data}

We use the primary CMB data from {\it Planck} 2018 \cite{2020A&A...641A...6P, 2020A&A...641A...1P}, including the 
temperature spectrum (TT) and polarization spectra (TE, EE) and lensing measurements. 

We also use a compilation of BAO data and RSD data from galaxy redshift surveys. The BAO and RSD data 
adopted in this paper is identical with Ref.~\cite{2020PhRvD.102d3517C}, including the measurements from 6dF 
Galaxy Survey (6dFGS) \cite{6dF-bao, 6df-rsd}, the Sloan Digital Sky Survey (SDSS) main galaxy 
sample (MGS) \cite{MGS-bao, MGS-rsd} and luminous red galaxy(LRG) 
\cite{LRG-bao, lrg-rsd}, SDSS DR12-BOSS \cite{dr12-combined}, SDSS DR14\cite{dr14-bao, dr14-rsd}, velocities from SNe 
\cite{sn-bao}, GAMA \cite{gama}, WiggleZ \cite{wigglez-rsd}, VIPERS \cite{vipers} and FastSound \cite{fastsound}.

\begin{table}
\begin{center}
\caption{\label{table:bao}Data points measured by BAO surveys used in this work.}
\begin{tabular}{cccc}
 \hline  \hline
Redshift & Measurement & Value & Surveys\\ \hline
0.106 & $r_{\rm s}/D_{\rm V}$ & $0.327 \pm 0.015$ & 6dFGS\\
0.15 & $D_{\rm V}/r_{\rm s}$ & $4.47 \pm 0.16$ & SDSS DR7-MGS\\
0.35 & $D_{\rm V}/r_{\rm s}$ & $9.11 \pm 0.33$ & SDSS DR7-LRG\\
0.38 & $D_{\rm M} (r_{\rm s,fid}/r_{\rm s})$ & $1518.4 \pm 22.4$ & SDSS DR12-BOSS\\
0.38 & $H(z) (r_{\rm s}/r_{\rm s,fid})$ & $81.51 \pm 1.91$ & SDSS DR12-BOSS\\
0.51 & $D_{\rm M} (r_{\rm s,fid}/r_{\rm s})$ & $1977.4 \pm 26.5$ & SDSS DR12-BOSS\\
0.51 & $H(z) (r_{\rm s}/r_{\rm s,fid})$ & $90.45 \pm 1.94$ & SDSS DR12-BOSS\\
0.61 & $D_{\rm M} (r_{\rm s,fid}/r_{\rm s})$ & $2283.2 \pm 31.9$ & SDSS DR12-BOSS\\
0.61 & $H(z) (r_{\rm s}/r_{\rm s,fid})$ & $97.26 \pm 2.09$ & SDSS DR12-BOSS\\
1.52 & $D_{\rm V}/r_{\rm s}$ & $26.005 \pm 0.995$ & SDSS DR14\\
\hline
\end{tabular}
\end{center}
\end{table}

\begin{figure} 
\includegraphics[width=0.5\textwidth]{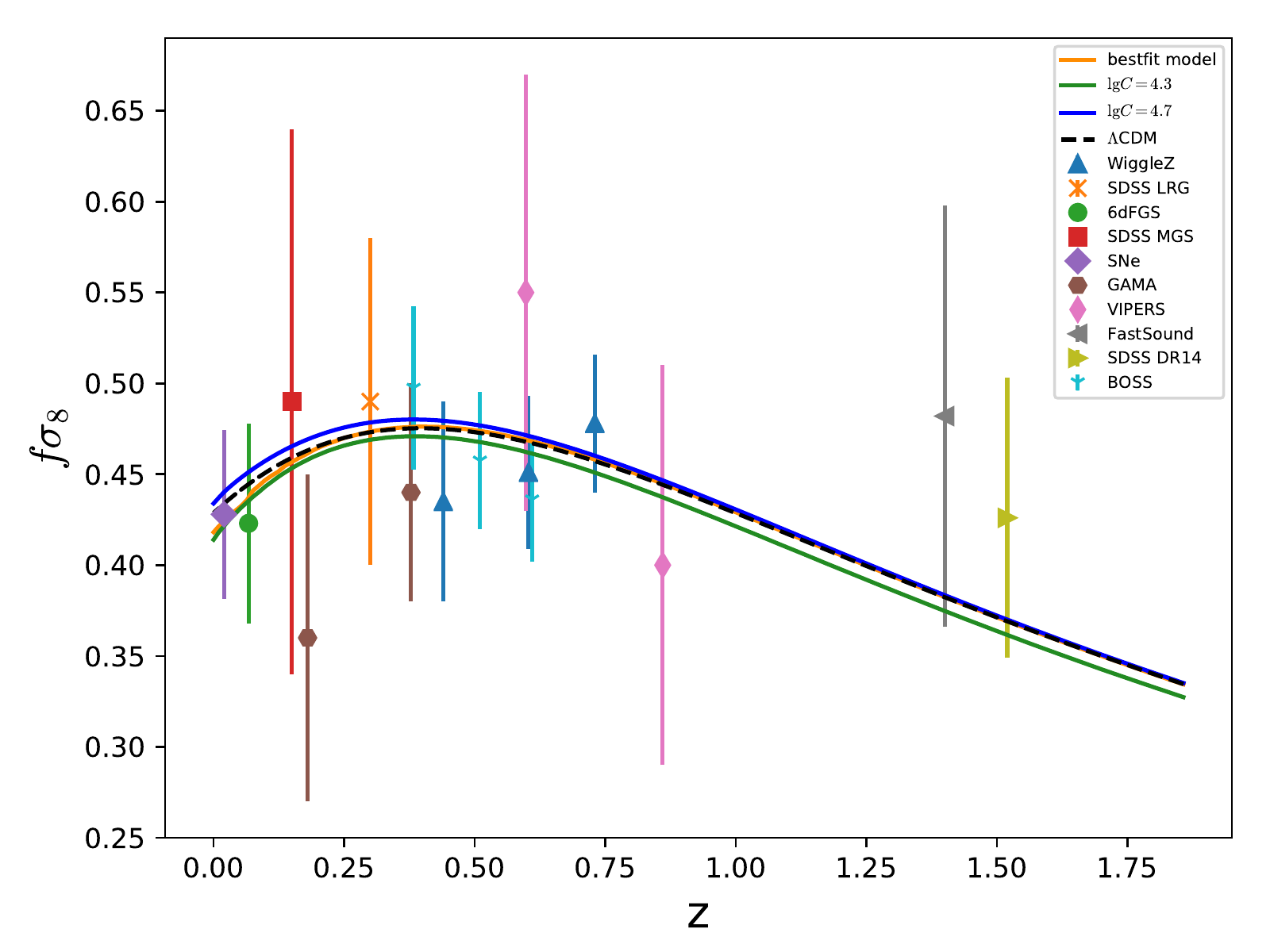}
\caption{The evolution of $f \sigma_{8}(z)$ with respect to the redshift $z$ in the $\Lambda$CDM model 
(dashed black line) and the same scalar-tensor models as Fig.~\ref{fig:back}.
The data points with error bars denote measurements from galaxy redshift surveys.}
\label{fig:rsd}
\end{figure}

Table \ref{table:bao} lists the BAO measurements used in this paper. 
$D_{\rm V}$ is a combination of the Hubble parameter $H(z)$ and the angular diameter distance $D_{\rm A}(z)$,
\be
D_{\rm V}(z) = \left[(1+z)^2D_{\rm A}^2(z)\frac{cz}{H(z)}\right]^{1/3},
\ee
while $r_s$ is the comoving sound horizon at the end of the baryon drag epoch. Note that the 
fiducial value of $r_s$ used by 
some observational groups is based on the analytical formula given by Eisenstein and Hu \cite{Eisenstein1998}, 
and it is essential to replace it by the accurate value given by {\tt CLASS}.

The RSD effect induced by the peculiar motion of galaxies can provide a powerful way to constrain the growth 
of structure. A large number of researches are conducted to measure the parameter combination 
$f(z) \sigma_{8}(z)$, in which the growth function is defined as 
\be
f(z) = \frac{{\rm d} \ln D}{{\rm d} \ln a},
\ee
where $D(a) = \delta(a)/\delta(a_0)$ is the linear growth function. We plot the current RSD constraints on the growth function in 
Fig. \ref{fig:rsd}. 
For the BOSS data, the measurements are $(1+z) D_{\rm A} (r_{\rm s,fid}/r_{\rm s})$, 
$H(z) (r_{\rm s}/r_{\rm s,fid})$ and $f \sigma_8$, and we use the full $9 \times 9$ covariance matrix to calculate the likelihood, 
so the correlation of $f \sigma_8$ among different redshift bins and the correlation
between $F_{\rm AP} (z) = (1+z) D_{\rm A} H(z)/c$ and $f \sigma_8$ are taken into account.
For the other RSD data, only the diagonal terms of the covariance matrix are used.
For the WiggleZ data, as noted in Ref. \cite{Planck2016-14}, the
points are conditionally plotted for the mean {\it Planck} cosmology according to the covariance matrix. 
In Ref. \cite{2017PhRvD..96b3542N}, the data points are rescaled by the ratios of $H(z) D_{A}(z)$ for the appropriate 
cosmology to take into account the Alcock-Paczynski (AP) effect. However, except BOSS and WiggleZ, the 
measurements at low redshift are almost independent of cosmology models (e.g. 6dFGS and velocities from SNe), 
or the error bars are too large for this to matter. The AP effect correction $\left[\cfrac{H(z) D_{\rm A}(z)}{H(z)^{\rm fid} D_{\rm A}(z)^{\rm fid}} - 1\right]$ is about $0.8 \%$, which is negligible compared with the RSD data error bar of  $17 \%$.
We have checked and found that the total impact of AP effect and the non-diagonal terms of the covariance matrix on the likelihood 
is about $\Delta \ln \mathcal{L} \sim 0.15$, so the correction due to AP effect is really small and 
the overall effect to the final constraints is negligible.
The best fit model preferred by CMB+BAO+RSD data set overlaps with the $\Lambda$CDM model at $z>0.25$.
As an example, we choose $\lg C = 4.3$ which is within the $1 \sigma$ bounds. We find that it fits the RSD data points
better than $\Lambda$CDM model, showing the ability of this model to provide a better fit to RSD data while being 
consistent with CMB+BAO data.

\begin{figure} [htb]
\includegraphics[width=0.4\textwidth]{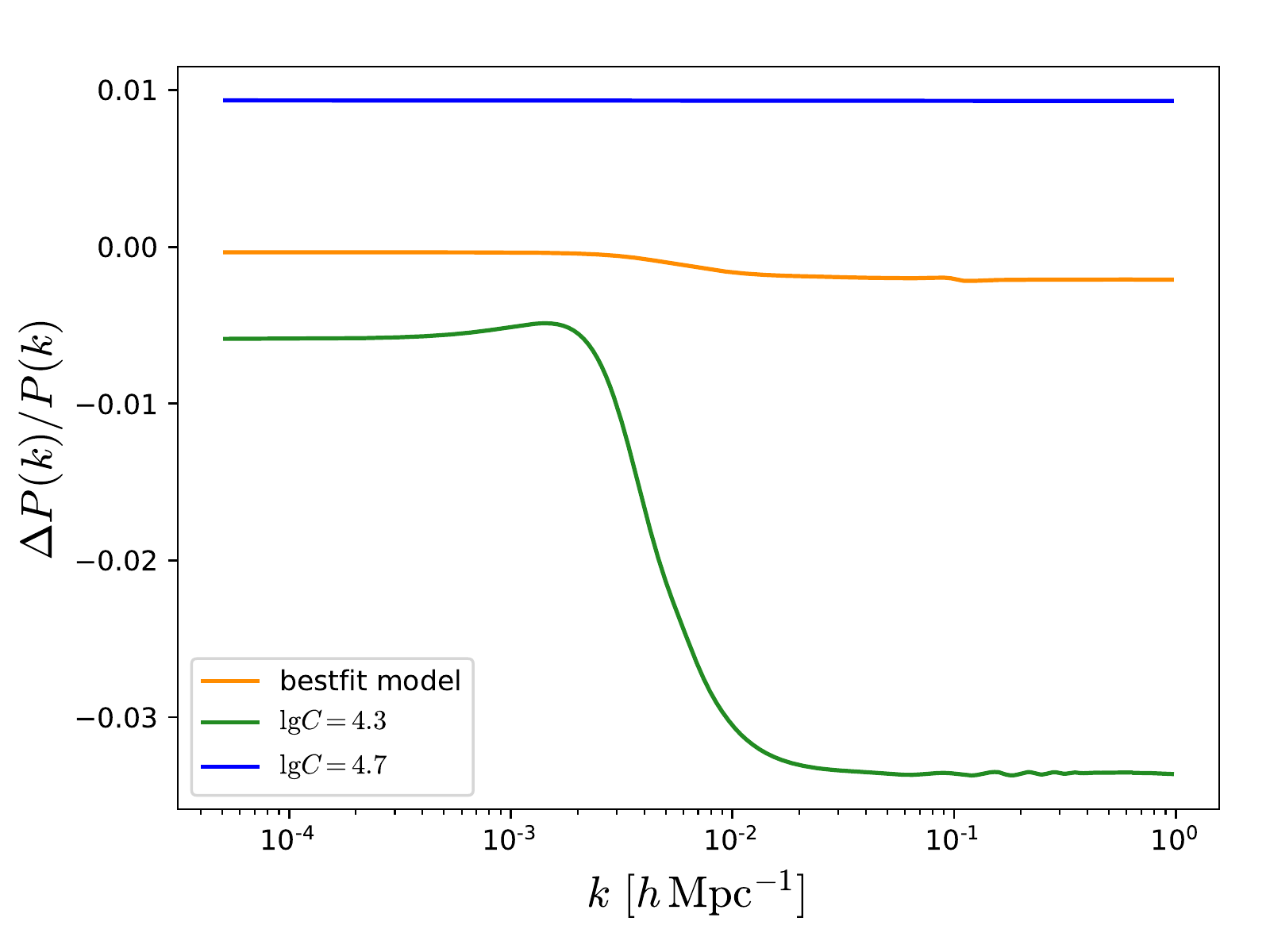}
\includegraphics[width=0.4\textwidth]{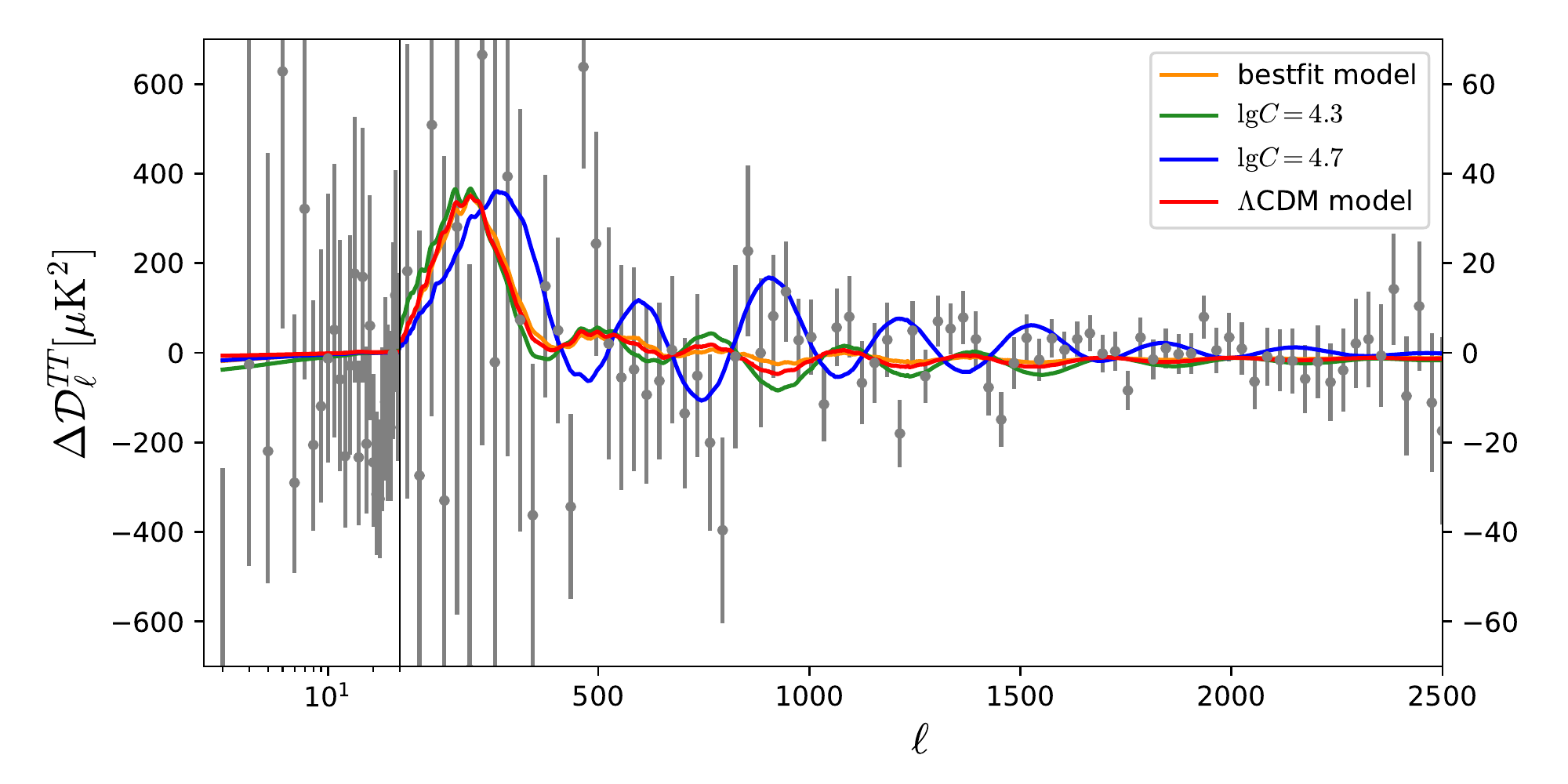}
\includegraphics[width=0.4\textwidth]{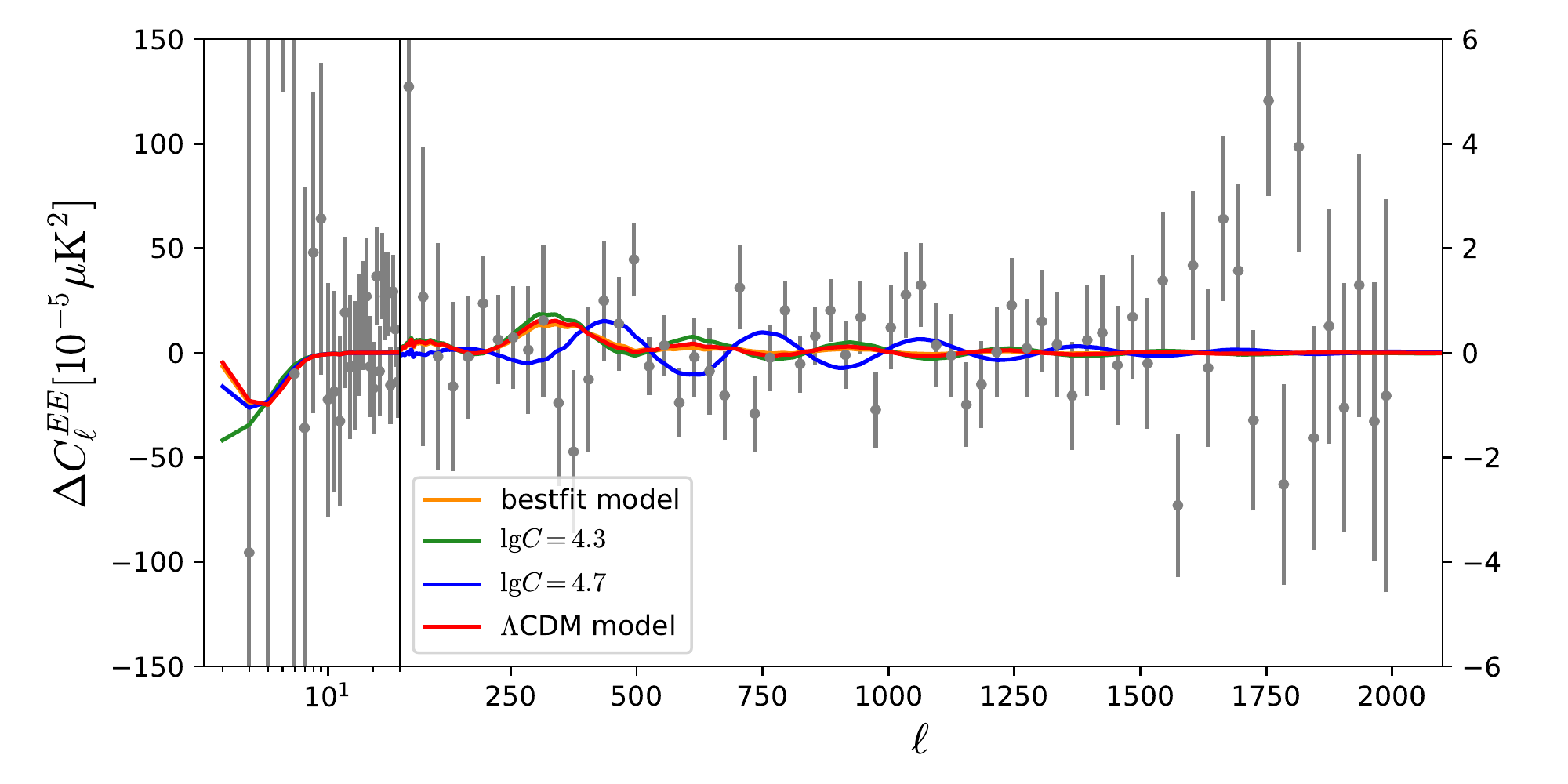}
\caption{
The relative difference of the matter power spectrum  (Top Panel), and the differences of the CMB temperature (Middle Panel) and $EE$ polarization (Bottom Panel) angular power spectrum between the three scalar-tensor models and the $\Lambda$CDM model. 
For the CMB, we also plotted the Planck 2018 best fit model and the binned Planck measurements \cite{planckurl}. 
In order to present the result more clearly, the horizontal axis switches from log to linear at $\ell = 30$ as in Ref.~\cite{2020A&A...641A...6P}, and we plot $\mathcal{D}_{\ell} = \frac{\ell (\ell +1)}{2\pi} C_{\ell}$ for TT spectrum and $C_{\ell}$ for EE spectrum.}
\label{fig:pk}
\end{figure}

\section{Results}

In Fig. \ref{fig:pk}, we show the relative difference of matter power spectrum and the CMB temperature (TT) and polarization (EE) angular power spectrum between $\Lambda$CDM model and the three example models we use. The locations of the acoustic peaks shift to larger or smaller angular scales, depending on the parameters used.
Note  for these four models the six cosmological parameters are set to be the same, so the  $\Lambda$CDM model (red line) adopts different cosmological parameters from the Planck 2018 best fit model.

\begin{figure*} 
\includegraphics[width=\textwidth]{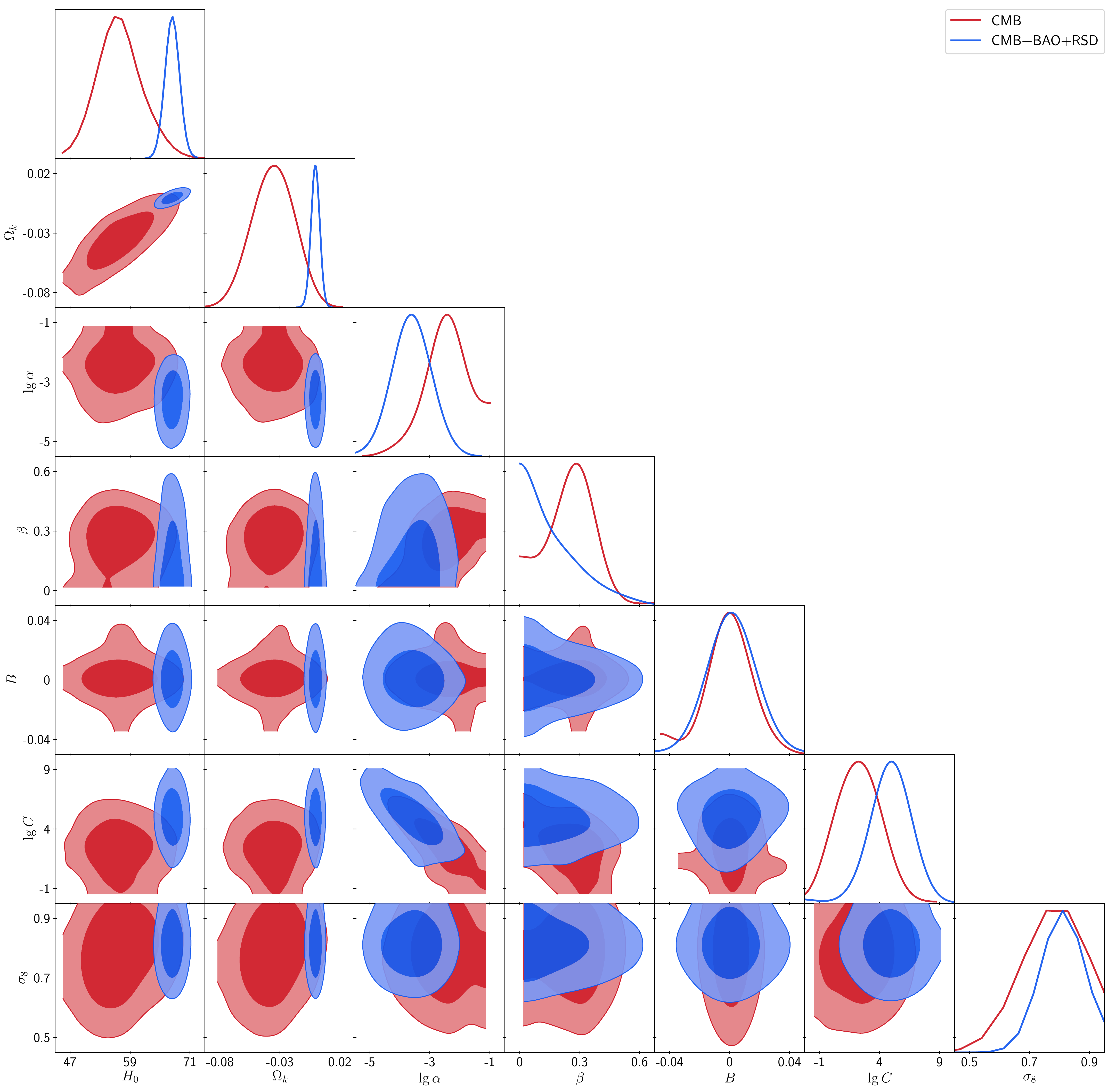}
\caption{The $68 \%$ and $95 \%$ confidence regions in the 
scalar-tensor model together with the 1d marginalized posterior distributions for the CMB, CMB+BAO+RSD data sets.
}
\label{fig:constraints}
\end{figure*}

We can derive constraints on single parameters or joint constraints on two parameters after marginalizing over the other parameters. 
Table \ref{table:constraints} lists the $1\sigma$ (68\%), $2\sigma$ (95.4\%) and $3\sigma$ (99.7\%) credible intervals for the various parameters. Two cases are given:  CMB only case,  in which  the  {\it Planck} 2018 temperature and polarizations data are used in the fitting,   
and the  CMB+BAO+RSD case, in which in addition to the CMB data, the {\it Planck} CMB lensing 
measurements and the BAO and RSD data set are also used. 
$A$ is not very well constrained by the CMB+BAO+RSD data set, so only $1 \sigma$ confidence region is given by MCMC.

To visualize the constraints from the observations, in Fig.~\ref{fig:constraints}, we also show  the 
$68 \%$ and $95 \%$  credible level contours of each pair of parameters, and  the 1-D marginalized posterior distributions of 
each parameter.  While the CMB data already constrained the parameters, 
the contours are tightened considerably with the additional BAO and RSD data, and the center of the contours are also 
shifted significantly. 
We found that although the best fit values of $H_0$ and $\sigma_8$
are nearly the same as the $\Lambda$CDM case, the corresponding errors are enhanced by a factor of $2-3$,
so in some sense, our model could potentially reduce the present tension in these parameters.
Specifically, for our model, the constraint from CMB+BAO+RSD data set is $H_0 = 67.5 \pm 1.2 ~{\rm km~ s^{-1} Mpc^{-1}}$,
while for $\Lambda$CDM model, $H_0 = 67.50 \pm 0.59 ~{\rm km~ s^{-1} Mpc^{-1}}$.
And the constraint from SN Ia and geometric distances from Milky Way parallaxes and eclipsing binaries is 
$H_0 = 73.5 \pm 1.4 ~{\rm km~ s^{-1} Mpc^{-1}}$ \cite{Reid2019}.
For the $\sigma_8$ tension, the constraint from CMB+BAO+RSD data set in our model is $S_8 \equiv \sigma_8 (\Omega_m/0.3)^{0.5} = 0.822 \pm 0.036$ and $\Omega_m = 0.311 \pm 0.011$,
while $\Lambda$CDM model yields $S_8 = 0.827 \pm 0.011$, $\Omega_m = 0.3123 \pm 0.0068$
and DES-Y1 galaxy clustering and weak lensing data gives $S_8 = 0.773^{+0.026}_{-0.020}$ and $\Omega_m = 0.267^{+0.030}_{-0.017}$ \cite{DESY1}.

\begin{figure}[h]
\includegraphics[width=0.35\textwidth]{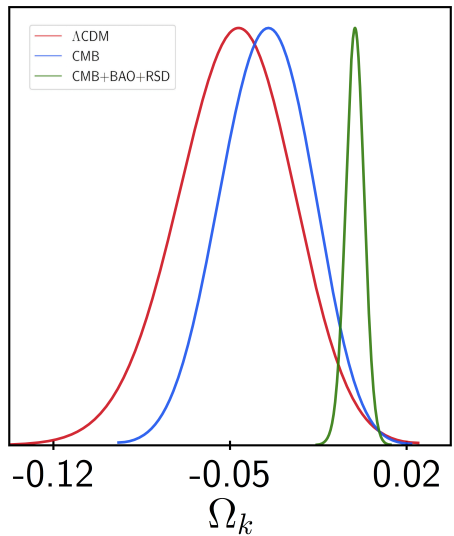}
\caption{The 1d marginalized posterior distributions of curvature. The red line represents the constraint from 
$\it Planck$ 2018 temperature and polarization data for the non-flat $\Lambda$CDM model. The blue line shows
the constraint from the same data but for our extended quintessence model. The green line shows the constraint on our model
with the CMB+BAO+RSD data.}
\label{fig:omega_k}
\end{figure}

Except for the parameter $\beta$, the 1-D posterior distributions we obtain all have narrow peaks well within the allowed range, so the 
result is not overtly dependent on the adopted prior.  The $\beta$ parameter distribution also has a peak, and for the CMB-only data 
it is in the middle of the allowed range, but for the CMB+BAO+RSD data combination, the  peak is  near 0, which is the 
 border of the prior. 
 
We find good consistency with the $\Lambda$CDM model for the parameters $\beta$, $A$ and $B$. 
Interestingly,  the fit gives non-zero value for $\alpha$, which deviates from the GR $\Lambda$CDM model limit ($\alpha=0$) 
at $ >3 \sigma$ level, but we should be very careful in interpreting this result. First, the corresponding $\omega$ value is quite large, 
so the deviation from the GR is in fact fairly small. Furthermore, 
if we inspect the value of the likelihood,  the maximum logarithm likelihood values are -1391.87 and -1391.90 
for our extended quintessence model and the GR $\Lambda$CDM model, respectively. This shows that 
 the GR $\Lambda$CDM model can fit the data almost as good as our extended quintessence model.  When only one parameter $\lg \alpha$
 is singled out, it may appear that the best fit  is several $\sigma$ away from the GR limit 
 (in our case, the minimum of $\lg \alpha$), but the Bayes estimator is obtained by integrating the posterior distribution over 
 the parameter space,  and this is not completely accurate for the GR $\Lambda$CDM model which essentially 
 has fewer parameters, i.e. lower dimensions. If we use the   the Akaike information criterion (AIC)  to compare our model with 
the GR $\Lambda$CDM model \cite{1974ITAC...19..716A, 2007PhRvD..76l3007G},
\be
\text{AIC} = -2 \ln \mathcal{L}_{\text{max}} + 2k,
\ee
where $k$ is the number of parameters, then the extended quintessence model is 
penalized for its additional four parameters, and the $\Lambda$CDM model would be the better model.

The data also favors a positive $C$, in which case the dark energy $V (\phi)$ is not constant but has an evolution.
There is  some degeneracy between the parameter $C$ and $\alpha$, as the $\alpha-C$ contour has a linear shape. This raises the possibility that the non-zero $\alpha$ best-fit may be partially due to this degeneracy, i.e. the observation data favors a dark energy model, and due to the degeneracy a non-zero $\alpha$ value is induced.    

We can derive the variation of the gravitational constant $G_{\rm rec}/G_0$ in this model, defined as the ratio of 
the gravitational constant during the Recombination Epoch and the present day. The $1 \sigma$ bound from CMB+BAO+RSD 
constraint is $0.97 < G_{\rm rec}/G_0 < 1.03$.

Finally, as we discussed earlier, in this work we consider non-flat geometry for both the extended quintessence model and the 
GR $\Lambda$CDM model. Fig.~\ref{fig:omega_k} shows the constraints on the curvature in our extended quintenssence model, 
compared with the $\Lambda$CDM model. As in $\Lambda$CDM model case, the closed model is also slightly favored 
by the {\it Planck} 2018 data in the scalar-tensor model. The flat universe case $\Omega_k = 0$ is still about $2 \sigma$ from the best fit. 
The peak of the distribution shifts toward the flat limit a little bit, from $-0.049$ to $-0.036$. So in this extended quintessence model the closed model is still a favored fit, only slightly closer to the flat case than the GR $\Lambda$CDM model. 

\begin{table*}
\begin{center}
\scriptsize
\caption{\label{table:constraints}CMB constraints and CMB+BAO+RSD joint constraints on the parameters.}
\begin{tabular}{ccccccc}
\hline  \hline
 & CMB & CMB & CMB & CMB+BAO+RSD &  CMB+BAO+RSD & CMB+BAO+RSD \\
Parameter & $68\%$ C.L. & $95.4\%$ C.L. & $99.7\%$ C.L. & $68\%$ C.L. & $95.4\%$ C.L. & $99.7\%$ C.L.  \\ \hline\\ \vspace{0.5em}
$\lg \alpha$ & $-2.41_{-0.65}^{+0.72}$ & $-2.4_{-1.3}^{+1.4}$ & $-2.4_{-2.2}^{+1.4}$ & $-3.59_{-0.54}^{+0.66}$ & $-3.6_{-1.2}^{+1.1}$ & $-3.6_{-1.8}^{+1.7}$  \\ \vspace{0.5em}
$\beta$ & $0.253_{-0.082}^{+0.13}$ & $0.25_{-0.25}^{+0.17}$ & $0.25_{-0.25}^{+0.29}$ & $0.16_{-0.16}^{+0.044}$ & $0.16_{-0.16}^{+0.29}$ & $0.16_{-0.16}^{+0.46}$ \\ \vspace{0.5em}
$\omega_0$ & $3.3 \times {10^4}~_{-3.2 \times 10^4}^{+6.3 \times 10^5}$ & $3.3 \times {10^4}~_{-3.3 \times 10^4}^{+1.3 \times 10^7}$ & $3.3 \times {10^4}~_{-3.3 \times 10^4}^{+8.3 \times 10^8}$ & $7.6 \times {10^6}~_{-7.2 \times 10^6}^{+8.3 \times 10^7}$ & $7.6 \times {10^6}~_{-7.5 \times 10^6}^{+1.9 \times 10^9}$ & $7.6 \times {10^6}~_{-7.6 \times 10^6}^{+3.0 \times 10^{10}}$ \\ \vspace{0.5em}
$10^7 A$ & $1.11_{-0.37}^{+0.62}$ & $1.11_{-1.0}^{+0.92}$ & $1.1_{-1.1}^{+1.2}$ & $1.993_{-0.046}^{+0.21}$ & $/$ & $/$  \\ \vspace{0.5em}
$B$ & $0.0006_{-0.0034}^{+0.0045}$ & $0.001_{-0.024}^{+0.025}$ & $0.001_{-0.034}^{+0.040}$ & $0.0011_{-0.0087}^{+0.0070}$ & $0.001_{-0.022}^{+0.025}$ & $0.001_{-0.050}^{+0.050}$  \\ \vspace{0.5em}
$\lg C$ & $2.2_{-1.7}^{+1.8}$ & $2.2_{-3.3}^{+3.2}$ & $2.2_{-3.8}^{+4.7}$ & $4.9_{-1.4}^{+1.4}$ & $4.9_{-2.8}^{+3.0}$ & $4.9_{-4.8}^{+4.1}$  \\ \vspace{0.5em}
$H_0 ~[{\rm km~ s^{-1} Mpc^{-1}}]$ & $56.9_{-4.8}^{+3.0}$ & $56.9_{-7.8}^{+8.7}$ & $57_{-10}^{+13}$ & $67.5_{-1.2}^{+1.2}$ & $67.5_{-2.5}^{+2.5}$ & $67.5_{-3.7}^{+4.1}$  \\ \vspace{0.5em}
$\Omega_{k}$ & $-0.036_{-0.015}^{+0.014}$ & $-0.036_{-0.030}^{+0.031}$ & $-0.036_{-0.043}^{+0.041}$ & $-0.0004_{-0.0029}^{+0.0031}$ & $-0.0004_{-0.0063}^{+0.0060}$ & $-0.0004_{-0.010}^{+0.0090}$  \\
\hline
\end{tabular}
\end{center}
\end{table*}

\section{Summary}
In this paper, we investigated a specific scalar-tensor theory,  with a quadratic scalar potential, i.e. an extended quintessence model 
of dark energy. We parameterize the Brans-Dicke parameter in a form similar to the harmonic attractor model, then 
we follow the standard formalism to derive the background and perturbation equations. 
Constraints on the model parameters are derived by using a MCMC program,  
with the latest cosmological data, including the {\it Planck} 2018 CMB data,  and the BAO and RSD data from various galaxy 
redshift surveys.

We found that in the quadratic potential extended quintessence model, the scalar field rolls down 
and oscillates around the minimum point of the potential $V (\phi)$.
This behavior is typical, for the oscillation is under-damped. And its energy density can be several orders of 
magnitude higher than the present day value, so it may play a role even in the early universe.

Our global fitting gives statistically significant non-zero $\alpha$ and $C$ values. This shows that at least the quadratic potential 
extended quintessence model considered here is allowed by current observations. However, we found that the maximum likelihood 
value is nearly the same for the GR $\Lambda$CDM model and our extended quintessence model. The non-zero value of $\alpha$ and $C$ is partly because the GR
$\Lambda$CDM  model has fewer dimensions, hence it is disadvantaged in the integration over parameter space. With less parameters, 
the GR $\Lambda$CDM model would be preferred with the Akaike Information Criterion.
The best fit values of $H_0$ and $\sigma_8$
are similar to those obtained in the $\Lambda$CDM case, however, the errors are enhanced by a factor of $2-3$.
So in some sense, our model could partially alleviate the present tension in these parameters.
The variation of the gravitational constant between the recombination epoch and the present day is constrained as $0.97 < G_{\rm rec}/G_0 < 1.03~ (1 \sigma)$.

In addition, we have constrained the curvature $\Omega_k$ in this gravity model with only CMB data. Compared with 
$\Lambda$CDM model, the peak of posterior distribution has a small shift from $-0.049$ to $-0.036$. This shows the 
possibility of alleviating the curvature problem raised recently in the context of modified gravity theories. 
\acknowledgments
We acknowledge the support of the Ministry of Science and Technology through grant 2018YFE0120800, the National Natural Science 
Foundation of China through grant No. 11633004, 11473044,  11773031, 11973047, and the Chinese Academy of Science grants QYZDJ-SSW-SLH017, XDB 23040100, XDA15020200. The computation of this work has been carried out  on TianHe-1(A) computer at the National Supercomputer Center in Tianjin, and the computers of the NAOC Astronomical Technology Center. 

\bibliographystyle{apsrev}
\bibliography{ms}

\end{document}